# Controlling elastic wave with isotropic transformation materials

**Zheng Chang[1], Jin Hu[2, a], Gengkai Hu[1, b], Ran Tao[2] and Yue Wang[2]**

1 *School of Aerospace Engineering, Beijing Institute of Technology, 100081,Beijing, China.*

2 *School of Information and Electronics, Beijing Institute of Technology, 100081, Beijing, China.*

**Abstract**: There are great demands to design functional devices with isotropic materials, however the transformation method usually leads to anisotropic material parameters difficult to be realized in practice. In this letter, we derive the isotropic transformed material parameters in case of elastodynamic under local conformal transformation, they are subsequently used to design a beam bender, a four-beam antenna and an approximate carpet cloak for elastic wave with isotropic materials, the simulation results validate the derived transformed material parameters. The obtained materials are isotropic and greatly simplify subsequent experimental implementation.

[a] Corresponding authors, bithj@bit.edu.cn
[b] Corresponding authors, hugeng@bit.edu.cn

Transformation method provides a direct way for finding material distribution when wave propagation pattern is prescribed, leading to many interesting devices for electromagnetic (EM) waves [1-4]. The method is then extended to acoustic wave for liquid materials since the Helmholtz's equation is also shown to have the transformation type solution [5, 6]. Recently, based on the interpretation of local affine transformation for a general space mapping, Hu et al. [7] also derive the transformed material parameters for elastic wave. Usually the transformed materials derived by transformation method are anisotropic with some unusual properties realized only with local resonant mechanism [4,8,9]. For practical purposes, especially for broadband applications, isotropic transformed materials are in great demand. Along this vein, Li and Pendry [10] proposed a quasi-conformal mapping to design a carpet cloak for EM waves, the transformed materials are quasi-isotropic, this greatly simplifies the experimental implementation [11,12]. Conformal mapping is proposed to design directional antennas for EM wave [13] and for acoustic wave [14], which leads to isotropic material parameters. In this letter, we will explore this possibility for elastic wave. Firstly the transformed relation for elastic wave under a local conformal mapping is derived, it is then applied to design a beam bender, a directional antenna with isotropic material, and to approximately design a carpet cloak in case of a quasi-conformal mapping.

We start by considering an elastodynamic problem in a virtual space described by Navier's equation

$$\nabla \cdot \boldsymbol{\sigma} = \rho \cdot \ddot{\mathbf{u}}, \quad \boldsymbol{\sigma} = \mathbf{C} : \nabla \mathbf{u}, \tag{1}$$

where $\mathbf{u}$ denotes displacement vector, $\boldsymbol{\sigma}$ is 2-order stress tensor, $\mathbf{C}$ is 4-order elasticity tensor and $\rho$ is the density. For a general spatial mapping $\mathbf{x}' = \mathbf{x}'(\mathbf{x})$ that transforms the virtual space to a physical space, Milton et al. [15] show that Navier's equation cannot keep its form. However, if we consider the transformation in a local point of view and adopt locally the affine transformation point by point, as commonly used to derive Navier's equation for elastodynamic problems, the transformed governing equation will still keep its form, i.e. Navier's equation. The transformed relation for elastic wave can then be obtained [7]. With the assumption of local affine transformation for the mapping, the deformation gradient tensor induced by the mapping $\mathbf{A} = \nabla_{\mathbf{x}} \mathbf{x}'$ can be decomposed as $\mathbf{A} = \mathbf{VR}$, where $\mathbf{R}$ and $\mathbf{V}$ denote a rigid rotation and a pure stretch tensor, respectively [16]. A local Cartesian frame $\mathbf{e}'$ at a point $\mathbf{x}'$ is established in the physical space, which is the principle frame of $\mathbf{V}$ ($\mathbf{V} = \lambda_1 \mathbf{e}'_1 \otimes \mathbf{e}'_1 + \lambda_2 \mathbf{e}'_2 \otimes \mathbf{e}'_2 + \lambda_3 \mathbf{e}'_3 \otimes \mathbf{e}'_3$). During the transformation, the stress, displacement, stiffness and density attached on a spatial element at the point $\mathbf{x}$ in the virtual space will experience a rigid rotation $\mathbf{R}$ and scaling along $\mathbf{e}'$ at the point $\mathbf{x}'$ in the physical space, symbolically as [7]

$$\mathbf{V_q R} : \mathbf{q} \mapsto \mathbf{q}', \quad \mathbf{q} = \boldsymbol{\sigma}, \mathbf{u}, \mathbf{C}, \boldsymbol{\rho}, \tag{2}$$

where $\mathbf{V_q}$ is the scaling tensor for quantity $\mathbf{q}$, and it has a diagonal form due to the specially established frame $\mathbf{e}'$, i.e.,

$$\mathbf{V}_{\boldsymbol{\sigma}} = \text{diag}[a_1, a_2, a_3], \mathbf{V_u} = \text{diag}[b_1, b_2, b_3], \mathbf{V_C} = \text{diag}[c_1, c_2, c_3], \mathbf{V}_{\boldsymbol{\rho}} = \text{diag}[d_1, d_2, d_3], \tag{3}$$

where $a_i$, $b_i$, $c_i$ and $d_i$ are scaling factors for stress, displacement, stiffness and density, respectively, remained to be determined. With local affine transformation, the form-invariance of Navier's equation and conservation of energy at each point lead to the following equations for determining the scaling factors [7]

$$\frac{a_i a_j}{d_J b_J} = \lambda_i, \quad \frac{a_i a_j}{c_I c_J c_K c_l b_k} = \frac{1}{\lambda_l}, \quad a_i a_j b_I = \frac{\lambda_j}{\lambda_1 \lambda_2 \lambda_3}. \tag{4}$$

The capital letter in index means the same value as its lower case but without summation.

Further, we consider a local conformal mapping at each point, i.e. $\lambda_1 = \lambda_2 = \lambda_3 \equiv \lambda$, then the scaling tensor $\mathbf{V_q}$ becomes isotropic, Eq. (4) becomes

$$\frac{a^2}{bd} = \lambda, \quad \frac{a^2}{c^4 b} = \frac{1}{\lambda}, \quad a^2 b = \frac{1}{\lambda^{N-1}}, \tag{5}$$

where $N = 2, 3$ for two dimensional and three dimensional problems, respectively. Obviously, there are non-unique solutions for the scaling factors $a$, $b$, $c$ and $d$ when $\lambda$ is given. However, we can for example express three of them in terms of the rest one and $\lambda$, leading to

$$b = \frac{1}{a^2 \lambda^{N-1}}, \quad c^4 = a^4 \lambda^N, \quad d = a^4 \lambda^{N-2}. \tag{6a}$$

Since the density $\rho$ and the modulus tensor $\mathbf{C}$ in the virtual space are isotropic, in the global frame, the transformed relations are finally derived as

$$\mathbf{C}' = a^4 \lambda^N \mathbf{C}, \quad \rho' = a^4 \lambda^{N-2} \rho, \quad \boldsymbol{\sigma}' = a^2 \mathbf{R} \boldsymbol{\sigma} \mathbf{R^T}, \quad \mathbf{u}' = a^{-2} \lambda^{1-N} \mathbf{R} \mathbf{u}. \tag{6b}$$

Both transformed modulus and density are isotropic. Equation (6b) provides in fact the transformed material parameters during a local conformal transformation, and they can be implemented into transformation method to design elastic devices, providing that the conformal transformation is guaranteed. We will demonstrate this through three examples: a beam bender, a directional antenna and an approximate carpet cloak.

A 2D beam bender can be designed by a mapping that transforms a rectangular plate into an arc shape. The conformal transformation can be obtained by setting the stretch along $\hat{\mathbf{r}}$ equals to that along $\hat{\boldsymbol{\theta}}$ direction at each point, since they are independent. This method has been used to design isotropic EM bender and the stretches are [17]

$$\lambda_r = \lambda_\theta = \lambda = \beta r / (ka), \tag{7}$$

where $\beta$ is the polar angle, $a$ is the original length of the rectangular plate and $k$ has an arbitrarily nonzero real value, as shown in Figs.1a and 1b. Different from EM wave, special caution must be taken to assure the impedance-matched condition between the transformed and the untransformed regions. For perpendicularly incident waves, the impedance-matched condition for both S and P waves can be expressed as [18]

$$\rho'\mathbf{C}' = \rho\mathbf{C} , \tag{8}$$

together with Eq. (6b), the impedance-matched transformed material parameters are given by

$$\mathbf{C}' = \lambda\mathbf{C}, \quad \rho' = \rho / \lambda , \tag{9a}$$

or

$$E' = \lambda E, \quad \upsilon' = \upsilon, \quad \rho' = \rho / \lambda . \tag{9b}$$

Noting that an isotropic material can be characterized by Young's modulus $E$ and Poisson's ratio $\upsilon$ as $C_{ijkl} = \dfrac{E\upsilon}{(1+\upsilon)(1-2\upsilon)}\delta_{ij}\delta_{kl} + \dfrac{E}{2(1+\upsilon)}(\delta_{ik}\delta_{jl} + \delta_{il}\delta_{jk})$ . To validate the transformed material parameters, a numerical validation is performed by the structural mechanics module of FEM software COMSOL Multiphysics, where a beam of a P-wave or S-wave is incident on the bender. The background is the structural steel with the material parameters $E = 200\,\mathrm{Gpa}$ , $\upsilon = 0.33$ and $\rho = 7850\,\mathrm{kg/m^3}$ . The simulation results on wave propagation pattern are shown in Figs.1c and 1d with frequency 20kHz and $\beta = \pi / 2$ , $a = 3$ , $k = \pi / 3$ for P-wave and S-wave, respectively, they clearly show that the waves are indeed guided to the new direction as desired.

The second example is to design a directional four-beam antenna for elastic wave with isotropic material. The antenna can be realized by Schwartz–Christoffel conformal transformation, defined by [13, 14]

$$w = \alpha \cdot 2i \cdot F[i \cdot \sinh^{-1}((i\frac{1+z}{1-z} - 1)^{-\frac{1}{2}})|2] + \gamma , \tag{10}$$

where $w = x' + iy'$, $z = x + iy$ and $F[\varphi | m]$ denotes the incomplete elliptic integral of the first kind with the modulus $m$, the coefficients $\alpha = -(2 + 2i)/[2K(-1) + i\sqrt{2}K(\frac{1}{2})]$, $\gamma = -1 + i$, where $K[\varphi]$ is the complete elliptic integral of the first kind. This transformation converts a cylindrical wave into plane one, once the transformation (shown in Figs.2a and 2b) is provided, the local principle stretch $\lambda$ can be evaluated, and the materials to realize this function are obtained by the transformed relations given by Eq. (9b). In numerical simulation, the conformal transformation is generated using MATLAB toolbox developed by Driscoll [19]. A line source of P-wave or S-wave is generated in the center of the antenna. The background is the same as that used in the beam bender. During the conformal transformation, singularities are formed at the intersection of the edges of the device, it is set to a finite large value in numerical computation, as in the references [13, 14]. Figures.2c and 2d show respectively the wave patterns for the generated P-wave and S-wave, again, a cylindrical wave is transformed into four plane wave beams, as expected. For acoustic wave, the stiffness tensor is characterized by only bulk modulus and the transformed material parameters given by Ren et al. [14] can be obtained as a special case.

Finally, we will show through an elastic carpet cloak that the isotropic transformed material can also be approximately applied to the case of quasi-conformal transformation. The carpet cloak is designed by the method proposed by Chang et al [20]. Since the stretch near the boundary of the cloak is very small, the impedance mismatch at the boundary is also small. In this case, we can further simplify the material parameter by assuming constant density or constant modulus. For constant density, we have $a^4 = 1/\lambda^{N-2}$, with help of Eq. (6b), the following transformed material parameters are derived

$$E' = \lambda^2 E, \quad \upsilon' = \upsilon, \quad \rho' = \rho. \tag{11a}$$

For constant modulus, we get

$$E' = E, \quad \upsilon' = \upsilon, \quad \rho' = \rho / \lambda^2. \tag{11b}$$

To validate the transformed material parameters, a carpet cloak is designed using Eq. (11a). The background is set to be the same as for the beam bender. Harmonic waves including both P-wave and S-wave are emitted from a line source, as shown in Fig. 3a. Figure 3c shows that through the isotropic carpet cloak, the scattering of the incident wave by a curvilinear boundary is the same as that by a linear boundary (Fig. 3a) if observed outside the cloak. Compared to the anisotropic elastic carpet cloak [7] shown in Fig.3d, the approximate isotropic carpet cloak works perfectly well. In our example, the anisotropy factor [10] is about 1.07. The same results can also be obtained for the carpet cloak designed with Eq. (11b). It is also worth to point out that the curvilinear boundary not only affects the reflection directions, but also the mode of the reflected waves due to the mode conversion at the boundary [18]. However, the designed carpet cloak can restore both the modes and the propagation directions.

To conclude, the isotropic transformed material parameters for elastic wave under local conformal mapping are derived, and they are used to design some interesting devices for elastic wave with isotropic materials, including a beam bender, a four-beam antenna. It is also shown through a carpet cloak that the isotropic transformed material parameter can also be applied to the case of quasi-conformal transformation.

**Acknowledgments**

This work was supported by the National Natural Science Foundation of China (10832002), and the National Basic Research Program of China (2006CB601204).

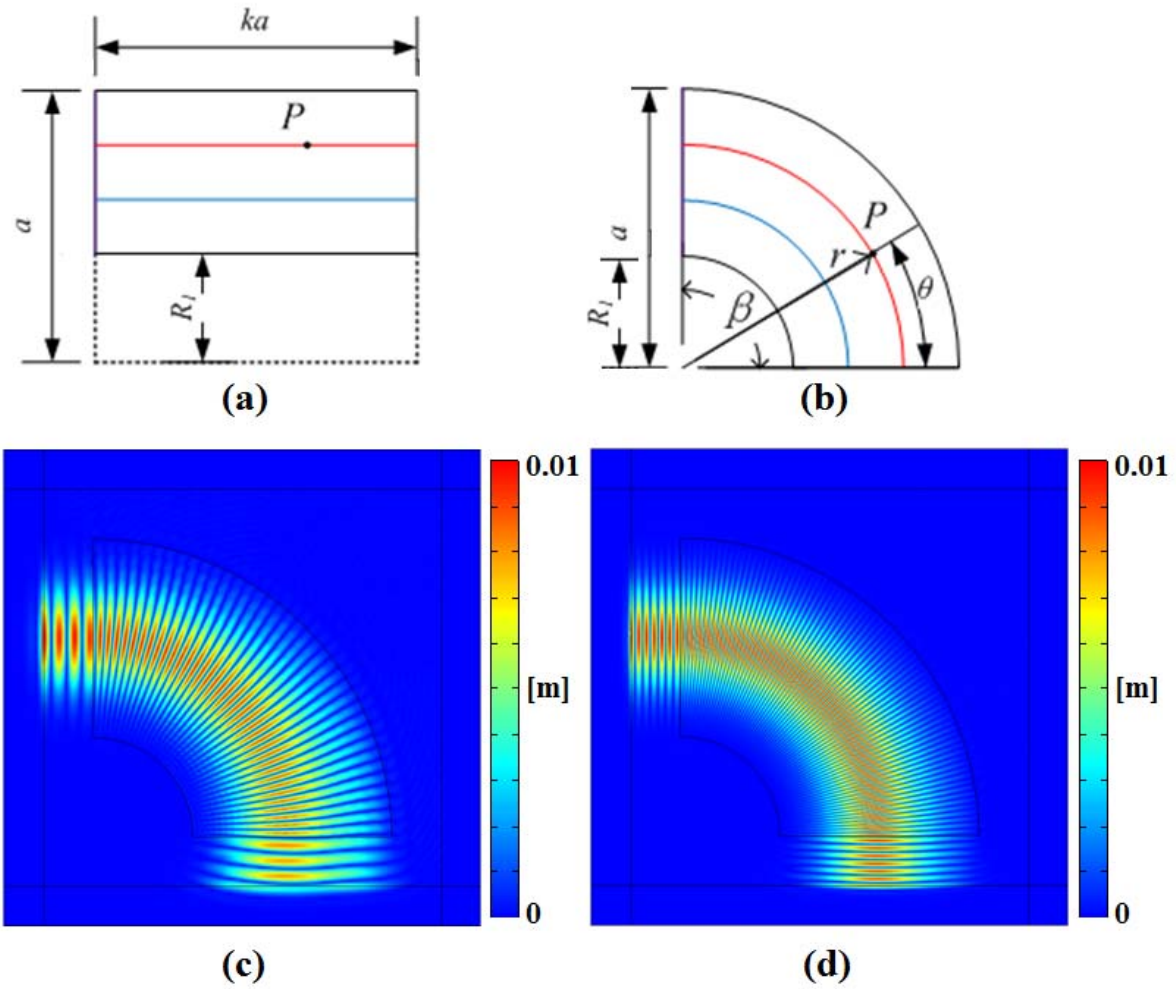

Figure 1 Transformation for a beam bender: (a) virtual space and (b) physical space. Total displacement $\sqrt{u_x^2 + u_y^2}$ for (c) P-wave and (d) S-wave in the bender.

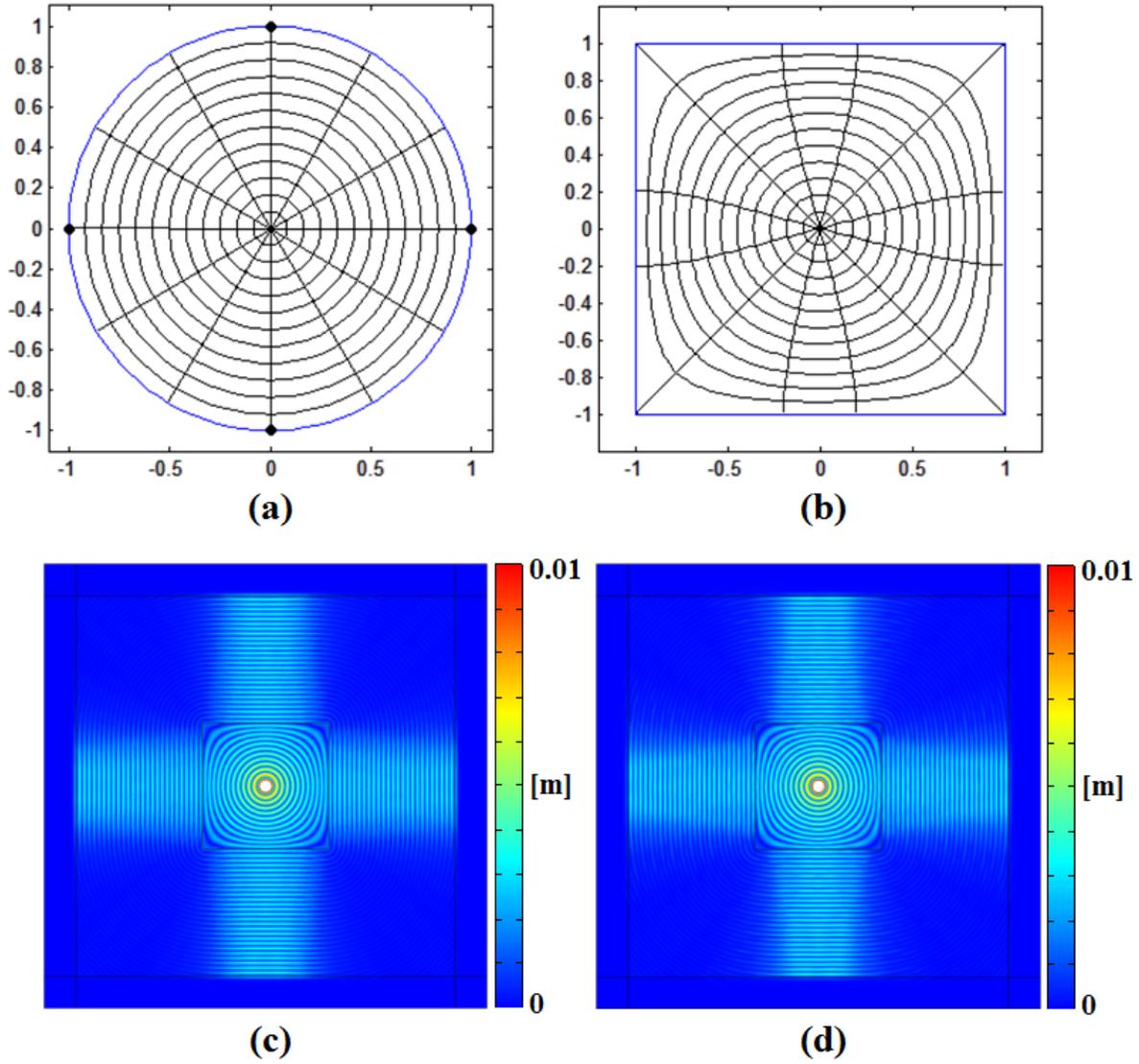

Figure 2 Transformation for the directional four-beam antenna: (a) virtual space $(\mathbf{x})$ and (b) physical space $(\mathbf{x}')$. Total displacement $\sqrt{u_x^2 + u_y^2}$ of the (c) P-wave and (d) S-wave incidents from the center of the directional four-beam antenna.

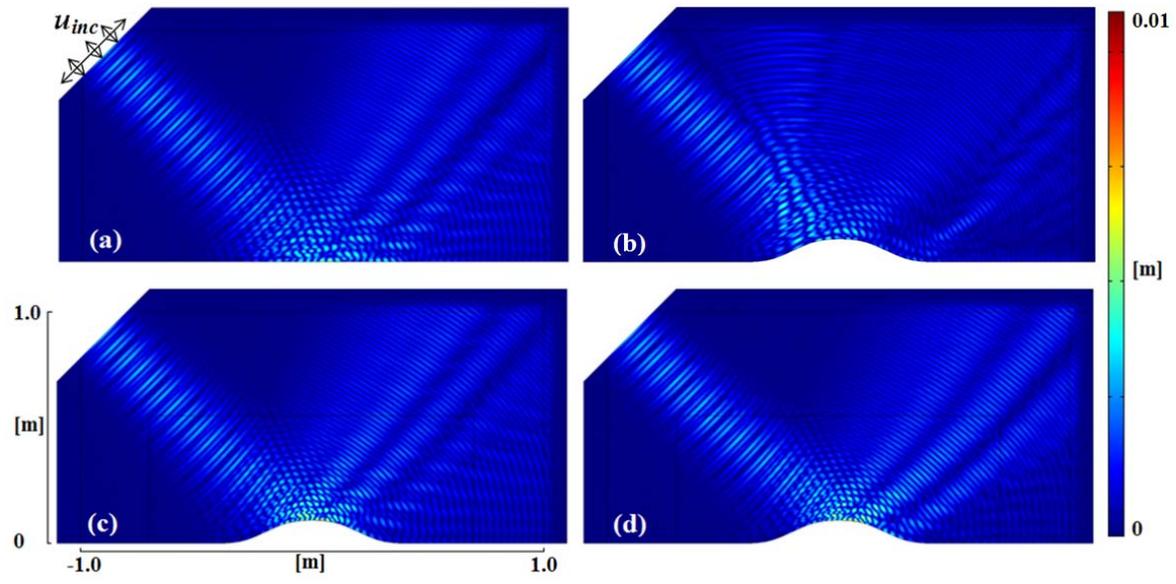

Figure 3 Wave (both P-wave and S-wave) incidents on the (a) linear and (b) curvilinear boundaries without the carpet cloak. Wave incidents on the curvilinear boundaries with the (c) isotropic and (d) anisotropic carpet cloak.